# Large-Area Aiming Synthesis of $WSe_2$ Monolayers

*Jing-Kai Huang, Jiang Pu, Chang-Lung Hsu, Ming-Hui Chiu, Zhen-Yu Juang, Yong-Huang Chang, Wen-Hao Chang, Yoshihiro Iwasa, Taishi Takenobu and Lain-Jong Li*

The $WO_3$ powders were placed in a ceramic boat located in the heating zone center of the furnace. The Se powders were placed in a separate ceramic boat at the upper stream side maintained at 270°C during the reaction. The sapphire substrates for growing $WSe_2$ were put at the downstream side, where the Se and $WO_3$ vapors were brought to the sapphire substrates by an $Ar/H_2$ flowing gas. The center heating zone was heated to 925°C. Note that the temperature of the sapphire substrates was at ~ 750 to 850°C when the center heating zone reaches 925°C. After reaching 925°C, the heating zone was kept for 15 minutes and the furnace was then naturally cooled down to room temperature. The morphology of the $WSe_2$ grown on the sapphire substrates varies with the substrate temperatures. The optical micrograph (OM) in Figure 1a shows that the $WSe_2$ grown at 850°C exhibits a triangular shape. Meanwhile, the sparsely distributed triangles indicate that the nucleation density of the $WSe_2$ is low. Figure 1b is the OM image for the $WSe_2$ grown at a lower temperature says at 750°C, where many small $WSe_2$ domains merge to form a continuous film.

..
Raman spectra for the monolayer and bilayer $WSe_2$ excited by a 473 nm laser are shown in Figure 2a, where the two characteristic peaks for monolayer $WSe_2$ at 248 $cm^{-1}$, assigned to $E^1_{2g}$ mode, and 259 $cm^{-1}$, assigned to $A_{1g}$ mode, are observed. Figure 2b compares the PL spectra (excited by a 532 nm laser) of monolayer and bilayer $WSe_2$ areas. The PL spectrum for the monolayer $WSe_2$ flake exhibits a strong emission at ~760 nm corresponding to the A excitonic absorption, whereas the emission intensity of the same peak is much lower in the bilayer $WSe_2$. All these PL observations further confirm the layer number assignment for our CVD $WSe_2$ samples.

The tunneling electron microscopy (TEM) image in Figure 3c shows the periodic atom arrangement of the $WSe_2$ monolayer, demonstrating that the CVD $WSe_2$ film is highly crystalline. The inset shows the selected area electron diffraction (SEAD) pattern taken with an aperture size (~160 nm) for the sample. Figure 3d is the magnified TEM image for the area squared by dashed lines in Figure 3c, where the high resolution TEM image and the corresponding SAED pattern with [001] zone axis (inset of figure 3c) reveal the hexagonal lattice structure with the lattice spacing of 0.38 and 0.33 nm assigned to the (100) and (110) planes.[49]

In conclusion, we have synthesized highly crystalline and large-area $WSe_2$ monolayers by the gas phase reaction of $WO_3$ and Se powders in a hot-wall CVD chamber. It is concluded that the hydrogen gas plays an important role to activate the reaction. Triangle and single crystalline $WSe_2$ monolayer flakes can be grown at 850°C, whereas a lower temperature 750°C results in a continuous and polycrystalline monolayer film.

## Experimental Section

**Characterizations**: The AFM images were performed in a Veeco Dimension-Icon system. Raman spectra were collected in a confocal Raman system (NT-MDT). The wavelength of laser is 473 nm (2.63eV), and the spot size of the laser beam is ~0.5μm and the spectral resolution is 3 $cm^{-1}$ (obtained with a 600 grooves/mm grating). The Si peak at 520 $cm^{-1}$ was used as a reference for wave number calibration. The $WSe_2$ films were transferred onto a copper grid for TEM observation. HRTEM imaging was performed on JEOL-2100F FEG-TEM operated at 100 kV. Chemical configurations are determined by X-ray photoelectron spectroscope (XPS, Phi V6000). XPS measurements were performed with an Mg Kα X-ray source on the samples. The energy calibrations were made against the C 1s peak to eliminate the charging of the sample during analysis. All electrical characterizations were performed using a semiconductor parameter analyzer (Agilent E5270) in a shield probe station inside an N2-filled glove box. We perform the impedance measurements using a frequency response analyzer (a Solartron 1252A frequency response analyzer with a Solartron 1296 dielectric interface controlled by ZPlot and ZView software).

[*]   J.-K. Huang, C.-L. Hsu, M.-H. Chiu, Dr. Z.-Y. Juang, Dr. Y.-H. Chang and Dr. L.-J. Li*
*Institute of Atomic and Molecular Sciences,
Academia Sinica, Taipei 10617, Taiwan*
Fax:(+886) 223668264
E-mail: (L.J.Li) lanceli@gate.sinica.edu.tw;

J. Pu and Dr. T. Takenobu*
*Department of Applied Physics, Waseda University, Tokyo 169-8555, Japan*
E-mail: (T.Takenobu)   takenobu@waseda.jp

Dr. W.-H. Chang
*Department of Electrophysics, National Chiao-Tung University, HsinChu, 300, Taiwan*

Dr. Y. Iwasa
*Department of Applied Physics, University of Tokyo, Tokyo 113-8656, Japan*

Dr. L.-J. Li*
*Department of Medical Research, China Medical University Hospital, Taichung, Taiwan*

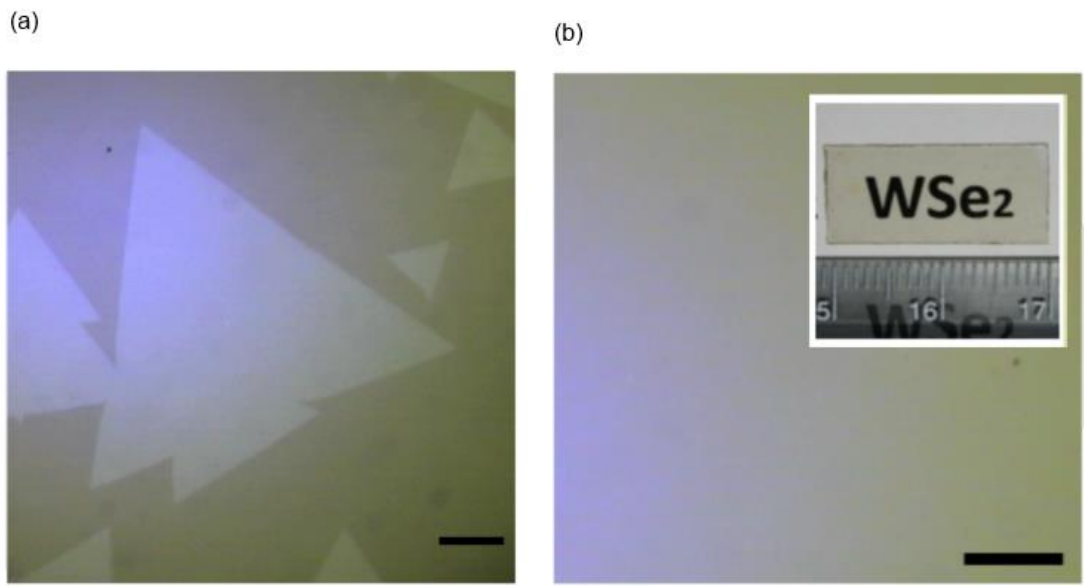

**Figure 1.** (a) The optical microscopy images of the WSe$_2$ monolayer flakes and monolayer film grown at 850 and (b) 750 °C respectively. Scale bar is 10 μm in length.

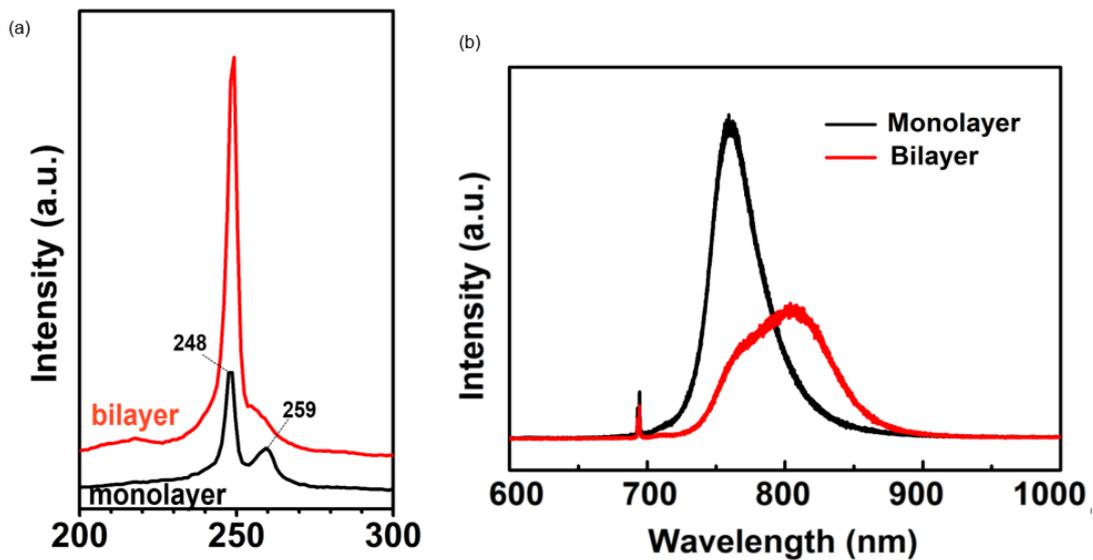

**Figure 2.** (a) Raman spectra for the monolayer and bilayer WSe$_2$, obtained in a confocal Raman spectrometer excited by a 473 nm laser. (b) Photoluminescence spectra for the CVD WSe$_2$ monolayer and bilayer, obtained in a microscopic PL system

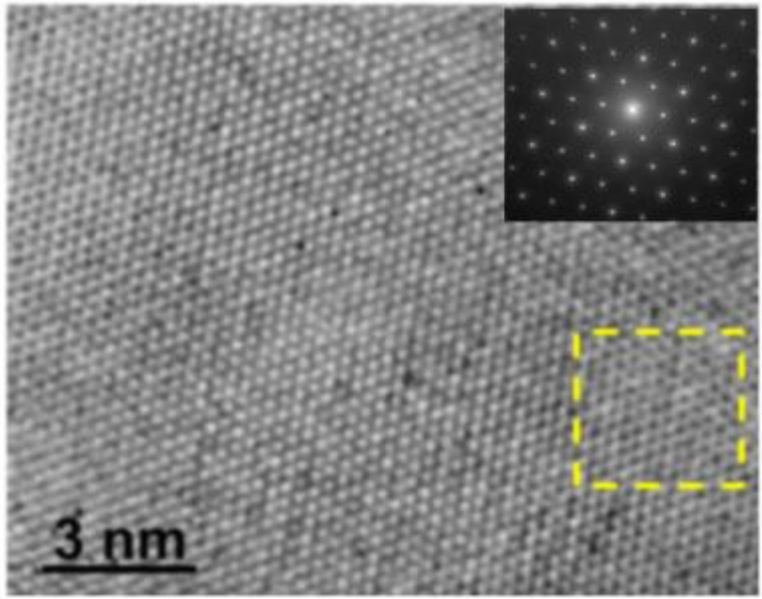

**Figure 3.** High resolution TEM image of WSe$_2$ monolayer with an inset showing its SAED pattern.